\newcommand{\vecr}{{\rm \bf r}}
\newcommand{\peps}{{\epsilon}}

\newcommand{\hp}{\hat{\rm \bf n}}
\newcommand{\hO}{{\hat{\rm \bf e}_{z}\,\!}}

\newcommand{\kg}{{k_{\rm g}}}

\newcommand{\mg}{{m_{\rm g}}}

\input{aipcheck}

\documentclass[
    ,final            
  ]
  {aipproc}

\layoutstyle{6x9}

\begin{document}

\title{Testing Gravity Theories In The Radiative Regime Using Pulsar Timing 
Arrays}


\author{K. J. Lee}{
	address={Max-Planck-Institut f\"ur Radioastronomie, Bonn 53121, Germany 
	(kjlee@mpifr-bonn.mpg.de)}
}

\begin{abstract}
General relativity has predicted the existence of gravitational waves (GW), 
which are waves of the distortions of space-time with two degrees of 
polarization and the propagation speed of light. Alternative theories predict
more polarizations, up to a maximum of six, and possible deviation of 
propagation speed from the light speed. The present paper reviews recent 
proposals to test the gravity theories in the radiation regime by observing GWs 
using pulsar timing arrays.

\end{abstract}

\maketitle

\section{Introduction}

Two important characteristics
of GWs are important to differentiate the validity of gravity theories in
the radiative regime; the polarization and dispersion of GW in vacuum. In
alternative metric theories, GW can have up to six possible polarization
states, four more then are allowed by GR, Furthermore, the propagation
speed of GW can deviate from the predication of GR that GW propagates
at light speed in vacuum, i.e. the effective graviton mass is zero.

Pulsar timing array is a unique technique to detect nano-Hertz GWs by timing 
multiple millisecond pulsars, which are very stable celestial clocks 
\citep{JHLM05}.  It turns out that a stochastic GW background leaves an 
angular-dependent correlation in pulsar timing residuals for widely spaced 
pulsars \citep{HD83}: the correlation $C(\theta)$ between timing residual of 
pulsar pairs is a function of angular separation $\theta$ between the pulsars.
One can analyse the timing residual and measure such a correlation to detect GWs 
\citep{JHLM05}. Lee et al.  (2008, 2010) have found that the exact form of 
$C(\theta)$ is very different from the one of GR, if the GW has extra 
polarization state or graviton mass is not zero.  Thus by measuring the 
correlation function, we can directly test gravity theories in the radiative 
regime.

\section{Pulsar Timing Correlation Functions and Gravity Tests}

A GW introduces extra signal in pulsar timing data. Let the unit vector of the 
GW propagation direction be $\hO$, GW frequency be $f$, the direction from the 
observer to the photon source (pulsar) be $\hp^{i}$.  The GW induced 
frequency-shift of a pulsar timing signal is \citep{LJR08, LJR10, BPPP08}
\begin{equation}\frac{\Delta\omega(t)}{\omega}=\frac{ \hp^{i}\hp^{j} }{2\left 
	(1+\frac{c}{2\pi f}{\rm \bf \kg} \cdot \hp\right )}  \left[ 
	h_{ij}(t,0)-h_{ij}(t-
	|\rm \bf D|/c,\rm \bf D )\right],\label{eq:z}\end{equation}
where the $\rm \bf D$ is the displacement vector from the observer to the 
pulsar, $ h_{ij}(t,0)$ and  $h_{ij}(t-
	|\rm \bf D|/c,\rm \bf D )$ are the metric perturbations by GW at the Earth and 
	at the pulsar when the received pulse was emitted, $\omega$ is the angular 
	frequency for the pulsar pulse, $f_{\rm cut}=\mg c^2/h$ is the cut-off 
	frequency of GW due to the graviton mass $\mg$, and the $\kg$ is the GW wave 
	vector given by \citep{LJR10}
	\begin{equation}
	{\bf \kg}(f) = \frac{2\pi \left(f^{2}-f_{\rm cut}^2\right)^{\frac{1}{2}}}{c}\, 
	\hO.
	\label{eq:dispf}
\end{equation}
The induced pulsar timing residuals $R(t)$ are given by the temporal integration 
of above the frequency shift at Earth given above, thus $R(t)=\int_{0}^{t} 
\frac{\Delta \omega(\tau)}{\omega} d\tau$.

The spatial metric perturbation $h_{ij}(t,\vecr)$ induced by a stochastic GW 
background is a superposition of monochromatic GWs with random phases and 
amplitudes.  It is \citep{Ma00} \begin{equation}
	h_{ij}(t,r^i)=\sum_{P=+,\times}\int_{-\infty}^{\infty}df\int d\Omega 
	h^{P}(f,\hO ) \peps^{P}_{ij}(\hO) e^{ i[2\pi f t-\kg^{i}(f) r_{i} ]},
	\label{eq:grbk}
\end{equation}
where $\Omega$ is the solid angle, index $i,j$ run from 1 to 3,
$h^{P}$ is the amplitude of the GW propagating in the direction 
of $\hO$ per unit solid angle per frequency of
polarization
state $P$, and the polarization tensor $\peps^{P}_{ab}$ of GWs are given in 
details in Lee et al. (2008).  The superscript $P$ takes value of `$+, \times$' 
for the two Einsteinian modes of GW polarization, `$b$' and `$l$' for the 
breathing and the longitudinal mode respectively, and `$sn, se$' for the shear 
modes. 

Such stochastic GW background leaves a correlation between timing residuals
of pulsars pairs \citep{HD83, LJR08}. Such correlation,
$C(\theta)$, depends on the angular distance $\theta$ between two pulsars as 
well as on the polarization of GW and graviton mass.

Lee et al. (2008) have calculated the pulsar timing correlation function for all 
the polarization modes of GW. For the Einsteinian modes and for the breathing 
mode, the cross-correlation function $C^{P}(\theta)$ is independent of 
earth-pulsar
distances and independent of the GW characteristic strain spectrum.  In 
contrast, for the modes that are not purely transverse,
the shear and longitudinal modes, the cross correlation functions
depend on the specifics of the strain spectra and on the pulsar
distribution in distance.

\begin{figure}[ht]
	\centering \includegraphics[totalheight=3in]{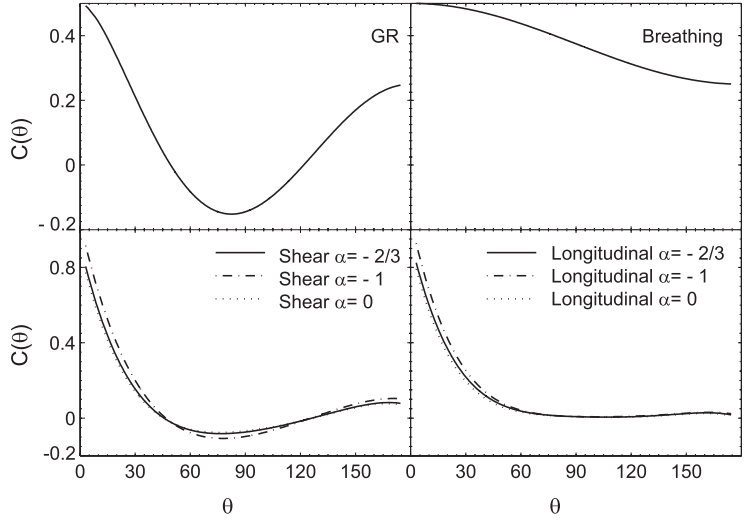}
        \caption{The normalized pulsar timing residual correlation
					coefficients. Here $\theta$
          is the angular separation between two pulsars. `GR' stands
          for the two transverse traceless modes, `$+$' and
					`$\times$'.  Results
          are given for several values of $\alpha$, the power-law
					index of the GW spectrum.}
	\label{fighdf}
\end{figure}

Fig.~\ref{fighdf} shows the correlation function according to different classes 
of GW polarization.
Clearly by comparison of these `theoretical' correlation curves with 
observations we can test the polarization state of GWs.  

\begin{figure*}[ht]
	\centering
		\includegraphics[totalheight=2.0in]{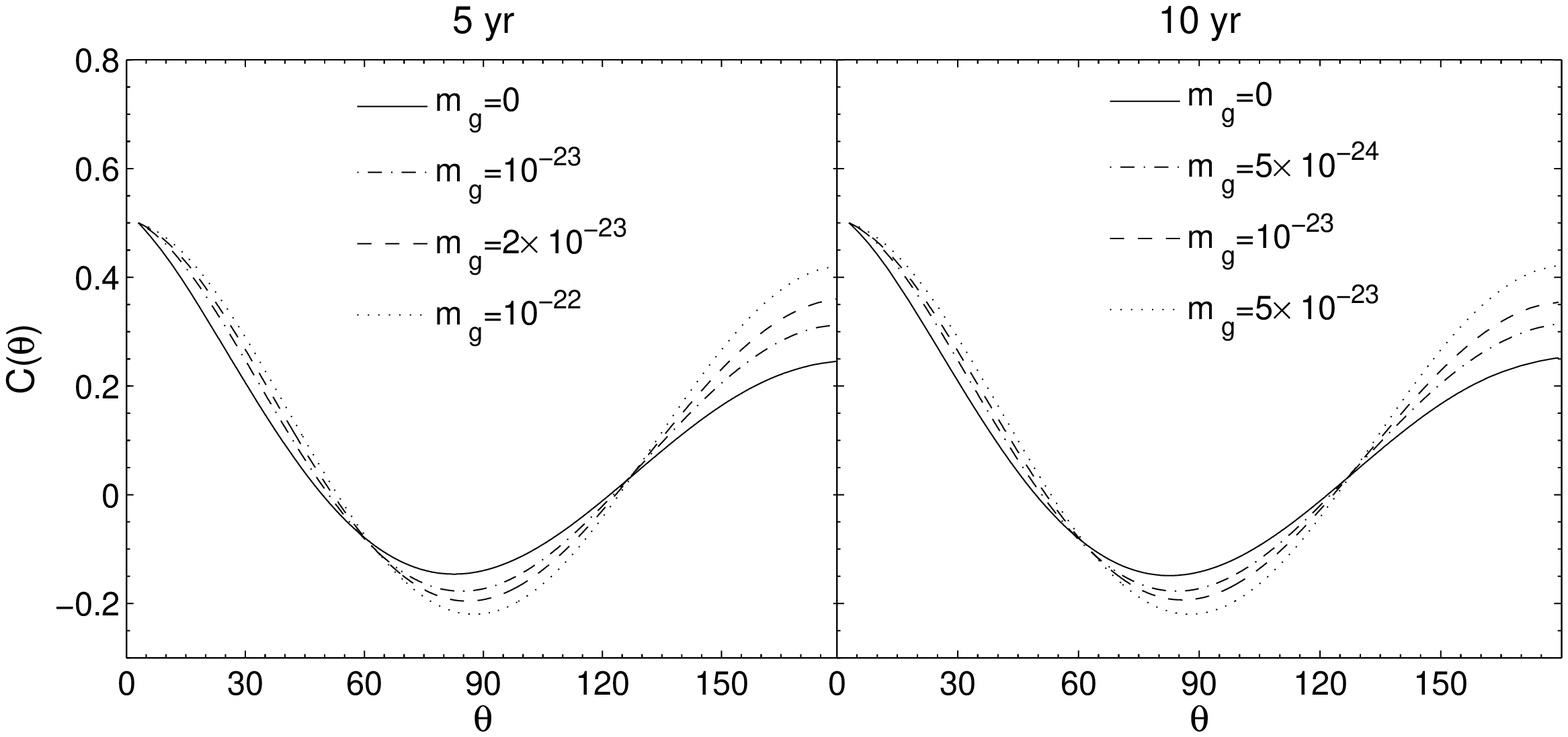}
	\caption{The atlas for cross-correlation functions
	$C(\theta)$. The label of each curve indicates the corresponding
	graviton mass in unit of electron-volts (eV). The left panel are
	the correlation functions for a 5-year bi-weekly observations.
	The right panel shows correlation functions for 10-year bi-weekly
	cases. We take $\alpha=-2/3$ for these results. These correlation are 
	normalized such that the $C(0)=0.5$
	for two \emph{different} pulsars.} \label{fig:hel}
\end{figure*}

Lee et al. (2010) have calculated the pulsar timing correlation function for a 
GW background with none-zero-mass graviton. They noted that the pulsar timing 
cross-correlation function for a massive GW background depends on the graviton 
mass, specific power spectra of the GW background, and on the observation 
schedule.  The 5-year and 10-year correlation functions are reproduced in 
Fig.~\ref{fig:hel}, where the graviton with the same mass introduces more 
deviation to the 10-year correlation function than it does to the 5-year one.  

Intuitively speaking, the necessary conditions for a positive detection of a 
graviton mass should be: 1.  The GW is strong enough such that the GW can be 
detected; 2. the physical effects of alternative theories should be strong 
enough to see the deviation from GR.  These intuition is confirmed by 
simulations in \citep{LJR08, LJR10}, which show that the high detection rate is 
achieved only if one has enough pulsar and if the graviton mass is large enough 
or if GW of alternative polarization modes is strong enough. 

For identifying the polarization modes, observation shows that if bi-weekly 
observations are made for five
years with RMS timing accuracy of 100\,ns, then 60 pulsars are required
for the longitudinal mode; 60 for the two spin-1 ``shear'' modes; and
40 for the spin 0 ``breathing'' mode and 40 pulsars are needed for the
detection of the GR modes.

For detecting massive graviton, simulations have shown that we need at least 60 
pulsars to be able to tell the difference between a massive GW background and a 
massless one.
For 5-year timing of 100 pulsar we can start to detect a graviton heavier than 
$2.5\times10^{-22}$ eV and we can achieve a limit of $m_{\rm g}=10^{-22}$ eV by 
using 5-year observation of 300 pulsars. We can achieve levels of $10^{-22}$ eV 
and $5\times 10^{-23}$ eV in a 10-year observation using 100 and 300 pulsars 
respectively.

\section{Conclusion and Discussion}
The stochastic GW background produces extra timing signals in pulsar TOA data, 
and one can detect the GW background by precise timing several pulsars and 
measuring the angular dependent correlations between the timing signals of 
several pulsars. A precise measurement of the angular
correlation function can, in principle, determine the GW polarization
properties of the GWs making up the stochastic background as well as the 
graviton mass.
A large number of pulsars with good timing precise of $\sim$ 100 ns are required 
to successfully perform tests for gravity theories in radiative regime.  In this 
regard, pulsar surveys' success in finding more millisecond pulsars is critical.  
To time such large mount of pulsars, the Large European Array for Pulsars 
\citep{BVK09}, the Five-hundred-meter Aperture Spherical Radio Telescope 
\citep{NWZZJG04, SLKMSJN09} and the Square Kilometer Array (SKA) will offer 
unique opportunities to detect the GW background and measure its properties. 

\section{Acknowledgment}

K.J.Lee gratefully acknowledges support from ERC Advanced
Grant ``LEAP'' Grant Agreement Number 227947 (PI: M.
Kramer), we also thank J. Verbiest for reading through the paper.

\bibliographystyle{aipproc}
\bibliography{ms}

\end{document}